\documentstyle[multicol,aps,prb,epsf]{revtex}
\begin{document}
\draft
\title{Two nonmagnetic impurities in the DSC and DDW state of
  the cuprate superconductors as a probe for the pseudogap}
\author{Brian M\o ller Andersen}
\address{\O rsted Laboratory, Niels Bohr Institute,
Universitetsparken 5, DK-2100 Copenhagen \O, Denmark}
\date{\today}
\maketitle
\begin{abstract}
The quantum interference between two nonmagnetic impurities
is studied numerically in both the d-wave superconducting (DSC) and the
d-density wave (DDW) state. In all calculations we include the
tunnelling through excited states from the CuO$_2$ planes to the
BiO layer probed by the STM tip. Compared to the single impurity
case, a systematic study of the modulations of the two-impurity local density of
states can distinguish between the DSC or DDW
states. This is important if the origin of the pseudogap phase is
caused by preformed pairs or DDW order. Furthermore, in the DSC state the study of the LDOS
around two nonmagnetic impurities provide further tests for the
potential scattering model versus more strongly correlated models.
\end{abstract}
\begin{multicols}{2}
\section{Introduction}
\noindent The study of magnetic and nonmagnetic impurities
in the CuO$_2$ planes of the High-T$_c$ superconductors is far
from settled. Experimentally, the local density of states (LDOS)
measured by scanning tunneling microscopy (STM) in Bi$_2$Sr$_2$CaCu$_2$O$_{8+\delta}$ (BSCCO) around a
nonmagnetic impurity such as Zn displays a sharp peak close to the
Fermi level on the impurity site and a second
maximum on the next-nearest neighbor
sites\cite{yazdani,pan}. Theoretically, the question remains
whether a traditional potential scattering formalism\cite{shiba,rosengren} or
more strongly correlated
models\cite{vojta} are needed to
describe the impurity effects. Though still a subject of controversy,
it was recently shown that at least for {\sl weak} impurities a potential
scattering scenario qualitatively agrees with the measured results for
optimally doped BSCCO\cite{hoffman,dhlee,zhangting,atkinson,capriotti}. Furthermore, it was
shown by Martin {\sl et al.}\cite{Martin} that both the
energetics and the spatial dependence of the resonance state 
around a strong potential scatterer (e.g. Zn) can be accounted for by
including the tunnelling (the filter) through excited states from the
CuO$_2$ planes to the top BiO layer probed by the STM tip\cite{hu}.
There is also evidence from nuclear magnetic resonance (NMR) measurements
that magnetic moments are induced around nonmagnetic
impurities\cite{NMRevidence}. In this paper we assume, however,
that the large potential scattering off the impurity site {\sl itself} is dominating the
final LDOS.\\
Recently the experimental ability to manipulate the positions of
surface impurities has increased the interest
in quantum interference phenomena between multiple impurities. This
includes the physics of quantum mirages\cite{manoharan} and 
various multiple impurity effects in
superconductors\cite{morrstavropoulos,hirschfeld,andersen,morrbalatsky}.
For example, it was shown in Ref. 17 that impurity interference can be
utilized as a sensitive probe for the gap symmetry of exotic superconductors. 
Motivated by the experimental progress we compare  
the expected LDOS around one and two strong nonmagnetic impurities in either the
d-wave superconducting (DSC) or the d-density wave (DDW) state. Though
still controversial we include the filter effect in all the
calculations presented below. As has become clear only
recently\cite{morrstavropoulos,hirschfeld,andersen}, we stress that the
probed impurities need be well separated (10-50 lattice constants) from
other possible defects.\\
The DDW state was recently proposed as a model for
the pseudo-gap state of the cuprates\cite{chakravarty}.
Any difference in the impurity modified LDOS between the DSC and DDW states may
reveal the hidden DDW order and distinguish between the scenario of preformed
pairs versus static staggered orbital currents as the 
origin for the pseudo-gap state\cite{Zhu,wang,morr}. Recently, there
has been several other proposals to probe the DDW order in the cuprates\cite{nguyen,franz}.\\
\section{Model}
In this section we briefly discuss the models for the DSC and DDW
states and how to calculate the LDOS around several impurities.
The BCS Greens function $\hat{G}^{0}\!({\mathbf{k}},i\omega_n)$ for the
unperturbed d-wave superconductor is given by
\begin{equation}
\hat{G}^{0}\!({\mathbf{k}},i\omega_n)=\left[
i\omega_n\hat{\tau}_0-\xi({\mathbf{k}})\hat{\tau}_3-\Delta({\mathbf{k}})\hat{\tau}_1 \right]^{-1},
\end{equation}
where $\hat{\tau}_\nu$ denotes the Pauli matrices in Nambu space,
$\hat{\tau}_0$ being the $2\times2$ identity matrix, $\xi(\mathbf{k})$
the quasi-particle dispersion, and $\omega_n$ is a Matsubara
frequency. For a system with d$_{x^2-y^2}$-wave pairing symmetry,
$\Delta({\mathbf{k}})=\frac{\Delta_0}{2} \left( \cos (k_x) - \cos
(k_y) \right)$.\\
In the DDW state the mean-field Hamiltonian is given by\cite{chakravarty}
\begin{equation}
H=\sum_{{\mathbf{k}}\sigma} \xi({\mathbf{k}})
c_{{\mathbf{k}}\sigma}^{\dagger}c_{{\mathbf{k}}\sigma} + i 
\sum_{\mathbf{k}\sigma} D({\mathbf{k}})
c_{\mathbf{k}\sigma}^{\dagger}c_{\mathbf{k+Q}\sigma}
\end{equation}
where $c_{\mathbf{k}\sigma}^{\dagger}$ creates an electron with
momentum $\mathbf{k}$ and spin $\sigma$, ${\mathbf{Q}}=(\pi,\pi)$ and $D({\mathbf{k}})=\frac{D_0}{2}
\left( \cos(k_x) - \cos(k_y) \right)$. Below, $\Delta_0=D_0=50\mbox{meV}$ and the lattice
constant is set to unity. The large value of the gap
corresponds roughly to the experimentally measured maximum gap in the underdoped regime of
BSCCO.\\
The Greens function for the clean DDW state is given by 
\begin{equation}\label{DDWGreens}
\hat{G}^0({\mathbf{k}},i\omega_n)\!=\!\frac{\left( \begin{array}{cc}
i\omega_{n} - \xi({\mathbf{k+Q}}) & -i D({\mathbf{k}}) \\
i D({\mathbf{k}}) & i \omega_{n} - \xi({\mathbf{k}})
\end{array} \right)
}{(i \omega_n-\xi({\mathbf{k}}))(i
\omega_n-\xi({\mathbf{k+Q}}))-D({\mathbf{k}})^2}.
\end{equation}
Performing the Fourier transform,
$\hat{G}^0({\mathbf{r}}_i,{\mathbf{r}}_j,i\omega_n)=\sum_{{\mathbf{k}}{\mathbf{k}}'}
\hat{G}^0({\mathbf{k}},{\mathbf{k}}',i\omega_n)
e^{i{\mathbf{k}}\cdot{\mathbf{r}}_i-i{\mathbf{k}}'\cdot{\mathbf{r}}_j}$,
of the Greens function with reference to the entries of
Eqn. (\ref{DDWGreens}) gives
\begin{eqnarray}\label{ddwrealspacegreen} \nonumber
\hat{G}^0({\mathbf{r}}_i,{\mathbf{r}}_j,i\omega_n)&\!=\!\!&\sum_{{\mathbf{k}}}[
G_{11}^0({\mathbf{k}},i\omega_n)\!+\! G_{12}^0({\mathbf{k}},i\omega_n)
e^{-i {\mathbf{Q}} \cdot{\mathbf{r}}_j}\!+ \\
G_{21}^0({\mathbf{k}},i\omega_n) e^{i {\mathbf{Q}} \cdot
{\mathbf{r}}_i}\!&+&\!G_{22}^0({\mathbf{k}},i\omega_n) e^{i
{\mathbf{Q}} \cdot ({\mathbf{r}}_i-{\mathbf{r}}_j)} ]
e^{i{\mathbf{k}}\cdot ({\mathbf{r}}_i-{\mathbf{r}}_j)},
\end{eqnarray}
with the sum extending over the reduced Brillouin zone.
The presence of scalar impurities is modelled by the
following delta-function potentials
\begin{equation}
\hat{H}^{int}\!=\! \sum_{\{i\}\sigma} U_i \hat{n}_{i\sigma},
\end{equation}
where $\hat{n}_{i\sigma}$ is the density operator on site $i$.
Here $\{i\}$ denotes the set of lattice sites hosting the
impurities and $U_i$ is the strength of the corresponding
effective potential. In this article all the presented results arise
from impurities modelled
by a potential, $U=-15t$, corresponding to -$4.5\mbox{eV}$. In
the DSC state this $U$ generates resonances at a few meV for a single
nonmagnetic impurity\cite{yazdani,pan,Martin}. The large scale of this 
potential renders the effects on the LDOS from corrections to other
energy scales around the impurity site less important. For instance, we have checked that
gap suppression near the impurity or slightly larger spatial
extension of the impurity does not qualitatively affect the 
results reported below. In general these effects tend to push the
resonances slightly further towards zero bias. We have also performed
calculations (not shown here) with other values of $U$ and comment on the results below.\\
The full Greens function
$\hat{G}({\mathbf{r}},\omega)$ in the presence of the impurities
can be obtained by solving the real-space Gorkov-Dyson equation
\begin{equation}\label{GorkovDyson}
\hat{G}(\omega) =
\hat{G}^{0}\!(\omega) \left(
\hat{I} -
\hat{H}^{int}
\hat{G}^{0}\!(\omega) \right)^{-1}.
\end{equation}
The size of the matrices in this equation depends on the number of impurities and the dimension of
the Nambu space. We have previously utilized this method to study
the electronic structure around impurities\cite{andersen} and vortices that
operate as pinning centers of surrounding
stripes\cite{andersenpinned}. This method is identical to the
traditional T-matrix formalism. However, for a numerical study of several
impurities at arbitrary positions we find it easier to solve
Eqn. \ref{GorkovDyson} directly.\\ 
The 2D Fourier transform of the clean Greens function
$\hat{G}^{0}({\mathbf{k}},\omega)$ is performed numerically by dividing the
first Brillouin zone into a $800\times800$ lattice and
introducing a quasi-particle energy broadening of $\delta = 1
\mbox{meV}$ with $\delta$ defined from the analytic continuation $i\omega_n\rightarrow\omega+i\delta$.
The differential tunnelling conductance is proportional to the
LDOS which is determined from the imaginary part of the full
Greens function.\\
So far nothing has been said about the form the
band-structure. It is still controversial which quasi-particle
energy applies to the DSC and DDW states\cite{morr,norman,shen}.
The expression for $\xi(\mathbf{k})$ is important
since it will influence the final LDOS around the
impurities. We illustrate this in the following by studying two generic
band structures: the nested situation, and a $t$-$t'$ band believed to
be relevant for BSCCO around $10\%$ hole doping. With the notation
$\xi(\mathbf{k})=\epsilon(\mathbf{k})-\mu$, and
\begin{equation}\label{disp}
\epsilon({\mathbf{k}})\!=\!-2t \left( \cos (k_x)\! +\! \cos (k_y) \right)
\!-\!4t' \cos (k_x) \cos (k_y),
\end{equation}
$t$ $(t')$ refers to the nearest (next-nearest) neighbor hopping
integral and $\mu$ is the chemical potential.  
The nested situation corresponds to $t'=\mu=0.0$ while the
parameters for the $10\%$ hole doped band are: $t=300 \mbox{meV}$,
$t'=-0.3t$ and $\mu=-0.9t$. These parameters correspond to the ones
previously studied for a single impurity by Morr\cite{morr}. As
discussed in Ref. 22 there are physical reasons to expect the nested
band to be relevant for the DDW state and the $t$-$t'$ band for the DSC
phase. However, recent photoemission measurements on LSCO by Zhou {\sl et al.}\cite{shen}
observed a Fermi surface consisting of straight lines connecting the
antinodal regions which may indicate that the nested band is more
relevant for impurity studies in LSCO. Thus we find it
important for study both cases below.\\
In the results presented we include the LDOS filter\cite{Martin}.     
This effect modifies the LDOS, $\rho({\mathbf{r}},\omega)=\sum_n
|\psi_n({\mathbf{r}})|^2 \delta(\omega - \epsilon_n)$, by including
the four nearest Cu neighbors in the underlying CuO$_2$ layer,
$\psi_n({\mathbf{r}})\longrightarrow\psi_n({\mathbf{r}}+{\mathbf{e}}_x)+\psi_n({\mathbf{r}}-{\mathbf{e}}_x)-
\psi_n({\mathbf{r}}+{\mathbf{e}}_y)-\psi_n({\mathbf{r}}-{\mathbf{e}}_y)$.
Here ${\mathbf{e}}_i$ denote the unit vectors in the CuO$_2$ plane. It
is important to keep in mind that the filtering effect is still
controversial. However, determining experimentally the interference effects around
two impurities in the DSC state may help resolve the relevance of the filter.
\section{Results}
\subsection{single impurity}
\begin{figure}
\centerline{\epsfxsize=\linewidth\epsfbox{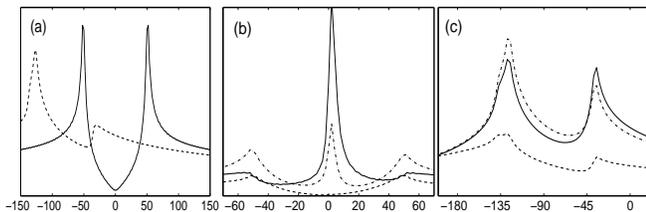}}
\caption{\label{1impDDW}DOS (arb. units) as a function of energy (meV) in the DDW
  state: (a) for the clean system with nested (solid) or $t$-$t'$
  (dashed) band. (b) DOS at ($0,0$) (solid), ($1,0$)
  (dashed), and ($1,1$) (dash-dotted) for a nested band with the impurity
  at ($0,0$). (c) same as (b) but for a $t$-$t'$ band.}
\end{figure}

\begin{figure}
\centerline{\epsfxsize=\linewidth\epsfbox{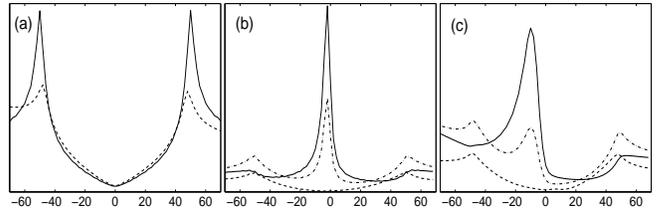}}
\caption{\label{1impDSC}Same as Fig. \ref{1impDDW} but for the DSC state.}
\end{figure}

Before studying the two impurity interference effects it is worthwhile
to briefly review the single impurity LDOS in the DSC and DDW states and discuss the influence of the
filter. Without the tunneling filter we find full agreement with 
previously published results\cite{dhlee,morrstavropoulos,Zhu,wang,morr}.
We will see that a single impurity is not a good probe for distinguishing between
these two states.\\ 
In the DDW phase one can utilize Eqn. (\ref{ddwrealspacegreen}) and (\ref{GorkovDyson}) to
calculate the full Greens function 
$\hat{G}({\mathbf{r}}_i,{\mathbf{r}}_j,i\omega_n)=\hat{G}^0({\mathbf{r}}_i-{\mathbf{r}}_j,i\omega_n)
+ \hat{G}^0({\mathbf{r}}_i,i\omega_n) T(i\omega_n)
\hat{G}^0(-{\mathbf{r}}_j,i\omega_n)$ with the T-matrix given by,
$T(i\omega_n)=U[1-U G^0(0,i\omega_n)]^{-1}$.
The single resonance condition, $1=U \mbox{Re}[G^0(0,\omega)]$, has
been previously studied for the DDW state without
the filtering effect\cite{Zhu,wang,morr}. It is well known that the
resulting LDOS strongly depends on the band structure.
In Fig. \ref{1impDDW}a we plot the DOS in the clean DDW state for the
nested and the $t$-$t'$ band without the filter. Even though the above resonance
condition is satisfied at certain energies for the
$t$-$t'$ band, we expect the large value of the DOS at all frequencies
to overdamp the impurity peaks. This is contrary to
the nested situation where a sharp impurity resonance is allowed to appear
in the gap. This is clearly verified in Fig. \ref{1impDDW}b(c) which depicts the LDOS
for the nested($t$-$t'$) set of band parameters including the
filter. The peaks in Fig. \ref{1impDDW}c
are not impurity resonances (note scale), which are overdamped, but
simply the shifted DDW gap edges. The
impurity can only slightly modify the amplitude of these gap edges. We
note that it is $t'$ which causes the impurity resonances to be
strongly overdamped. When $t'=0$, $\mu \neq 0$ the density of states always vanishes at minus
the chemical potential\cite{wang} allowing a well-defined resonance
peak to appear.\\ 
As is evident from Fig. \ref{1impDDW}b the most important influence of the filter is to
shift the LDOS maximum from the nearest neighbors to the  
impurity site and induce a second maximum on the next-nearest
neighbor sites\cite{comment1}. This weight redistribution is identical to the
situation in the superconducting phase\cite{Martin}.\\
In the DSC state, the clean DOS is plotted in
Fig. \ref{1impDSC}a for both the nested and the $t$-$t'$ band. By
comparison to Fig. \ref{1impDDW}a we see the well known result that
the nested DOS is identical for the clean DDW and DSC phase. Indeed this motivated
the original studies of single impurity resonances in the DSC versus
DDW states\cite{Zhu,wang,morr}. The single impurity resonance
condition in the DSC phase, $1=U\mbox{Re}[G^0(0,\pm \omega)]$, generates peaks at 
positive {\sl and} negative energies around a single nonmagnetic impurity.  
However, the majority of the quasi-particle weight may reside on only one of these
resonances\cite{andersen}.     
\noindent It is evident from {\sl both}
Fig. \ref{1impDSC}b and Fig. \ref{1impDSC}c
that indeed only one resonance has weight. This is contrary to the
situation without the filter\cite{morr}. Thus by comparing  
Fig. \ref{1impDDW}b to Fig. \ref{1impDSC}b (or \ref{1impDSC}c) the result is two almost identical
figures. Therefore, since no qualitative difference is
guaranteed to exist the single nonmagentic impurity cannot easily
distinguish the DSC and DDW phases . However, as shown below, the interference
between several impurities can be utilized to
{\sl tune the amplitude} of the potential resonances and thus clearly
distinguish the phases.\\
The impurity LDOS plotted in Fig. \ref{1impDDW} and
Fig. \ref{1impDSC} was for $U=-15t$. Though of less experimental
relevance, we briefly mention another difference between the DSC and
DDW states. This relates to the fate of the resonance in the unitary limit, $U\rightarrow\infty$:
for the DDW phase the resonance energy approaches minus the chemical potential,
$\omega=-\mu$, whereas it approaches the Fermi level in the DSC phase
(except for a small residual energy shift caused by a possible
particle-hole asymmetry\cite{joynt}). The different resonance energy (as $U\rightarrow\infty$) arises from the
way the chemical potential enters the bands of the clean DDW
($E_\pm({\mathbf{k}})=|\sqrt{\epsilon({\mathbf{k}})^2+D({\mathbf{k}})^2}
\pm \mu|$) and DSC ($E_\pm({\mathbf{k}})=\sqrt{(|\epsilon({\mathbf{k}})| \pm
  \mu)^2+\Delta({\mathbf{k}})^2}$)
states\cite{Zhu,wang}.\\   
\subsection{two impurities, nested band}
\begin{figure}
\centerline{\epsfxsize=\linewidth\epsfbox{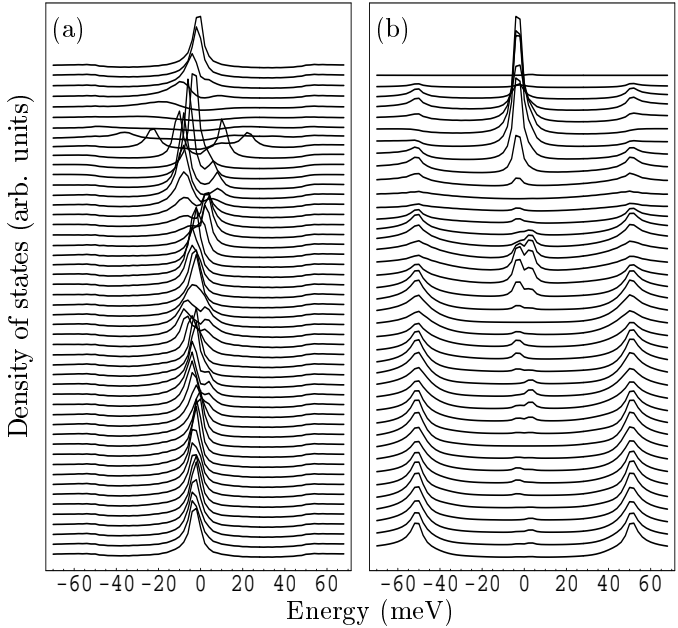}}
\caption{\label{DSCnested}(a) DOS at $(0,0)$ as a function of the distance between
  two nonmagnetic impurities (here: $t'=\mu=0.0$). One impurity is fixed at
  $(0,0)$ while the other moves from $(0,0)$ (top) to
  $(10,0)$ (bottom). (b) the impurities are fixed
  at ($\pm 1,0$) while the STM tip is moved from ($0,0$) (top)
  to ($8,0$) (bottom). The difference between each scan is 0.2 lattice
  constants and the graphs are off-set for clarity.}
\end{figure}

In general when several impurities are in close proximity the
resonances split, and one expects to see additional peaks in the
density of states. The evolution of the LDOS as a function of distance and angular
orientation between two nonmagnetic impurities in the DSC state
has been already studied by several authors\cite{morrstavropoulos,hirschfeld,andersen}. In the following
we elaborate on this work by a numerical study of the 
LDOS including the filtering effect and study for the first time the
quantum interference between two strong nonmagnetic impurities in the
DDW state. In the superconducting phase Fig. \ref{DSCnested}a shows the resulting
LDOS for the nested band when one impurity is fixed at the origin $(0,0)$ while the other is moved out
along a crystal axis to $(10,0)$. In Fig. \ref{DSCnested}b the impurities
are fixed at $(-1,0)$ and  $(+1,0)$ while the STM tip is moved from
$(0,0)$ to $(8,0)$. As seen from both figures there are strong
variations in the LDOS in agreement with previous studies without
the extra tunnelling effect\cite{morrstavropoulos,andersen}.  
\begin{figure}
\centerline{\epsfxsize=\linewidth\epsfbox{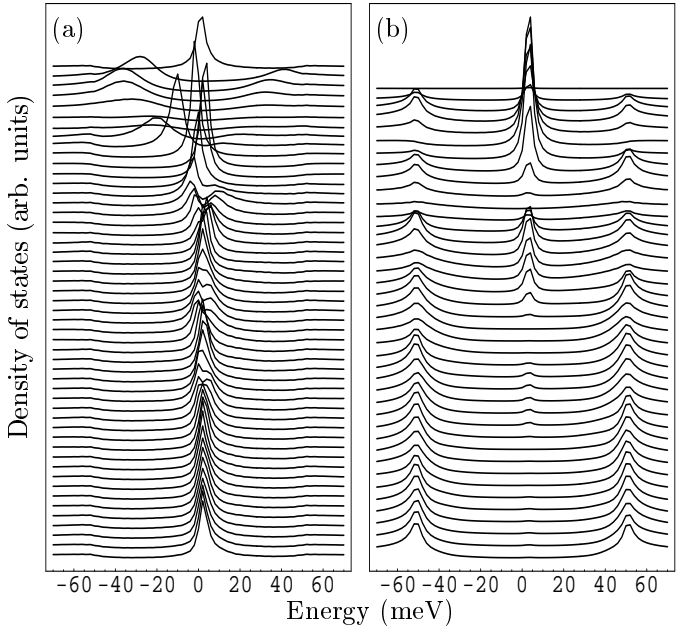}}
\caption{\label{DDWnested}Same as Fig.
\ref{DSCnested} but for the d-density wave state.}
\end{figure}
\noindent The number of apparent resonances, their energetic position and width
strongly depend on the impurity configuration and the position of
the STM tip. In particular, for certain impurity separations the
resonances completely disappear. In Fig. \ref{DDWnested} we show the LDOS for the same impurity and
STM positions as in Fig. \ref{DSCnested} but for the DDW
state. Clearly, strong quantum interference between the two 
nonmagnetic impurities also exists in this state. However, by comparison with Fig
\ref{DSCnested} it is evident that the additional resonance states in
the DSC allows one to distinguish this from the DDW phase. We have
performed identical calculations to the ones presented in 
Fig. \ref{DSCnested}-\ref{DDWnested} for other (but still large) values of the
scattering potential $U$, and always find qualitatively the same
interference pattern.\\
As mentioned above, the resonances split when two
impurities are in close proximity. It is therefore nontrivial that
only a single, nondispersive peak is seen in
e.g. Fig. \ref{DDWnested}b. This is closely
connected to the particular STM scan and one may worry about
the robustness of this result. However, we always find that whenever
the impurity positions are invariant under mirror reflection through
the STM scan line, only a single nondispersive peak
remains\cite{comment2} in the DDW state. Importantly, for these same
configurations we find the alternating double peak structure (similar
to Fig. \ref{DSCnested}b,\ref{DSCnotnested}b) to be a robust
feature in the superconducting phase. Furthermore, as expected for a d-wave gap\cite{andersen}, we
find (not shown) that the quantum interference patterns are longer ranged along
the nodal directions than along the Cu-O bonds.\\ 
As expected from the discussion of the single impurity in the DDW
state, we end this section by noting that when $t'=0.0$, $\mu \neq 0.0$
the interference pattern is identical to that shown in
Fig. \ref{DDWnested} except for a shifted (by $-\mu$) energy range.
\subsection{two impurities, $t$-$t'$ band}
\begin{figure}
\centerline{\epsfxsize=\linewidth\epsfbox{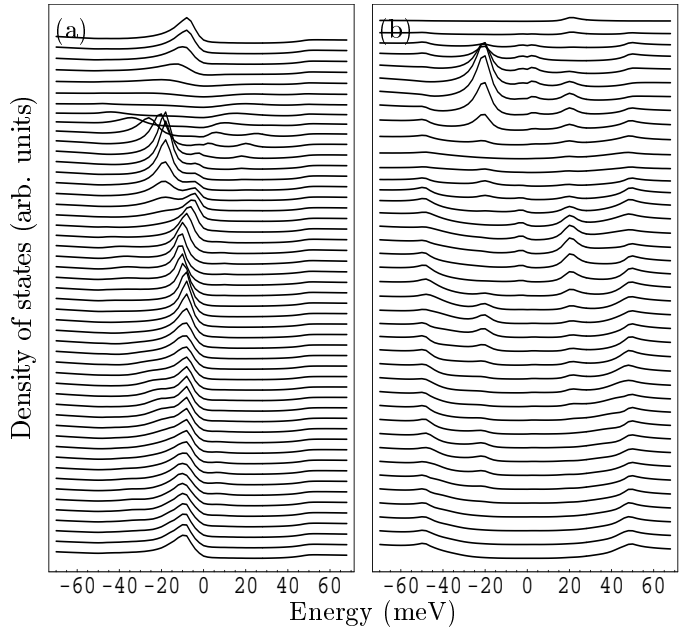}}
\caption{\label{DSCnotnested}Same as Fig. \ref{DSCnested} but for $t'=-0.3t$
  and $\mu=-0.9t$.}
\end{figure}

We now turn to the quasi-particle dispersion given by Eqn. (\ref{disp})
with $t'=-0.3t$, $\mu=-0.9t$. In this case we know from
Fig. \ref{1impDDW}a and Fig. \ref{1impDSC}a that the clean DOS is 
clearly different in the DDW and DSC states. This section serves as an
illustration of the importance of the quasi-particle dispersion in the
final LDOS. Fig. \ref{DSCnotnested}
shows the LDOS in the superconducting phase 
from the same STM and impurity positions as Fig. \ref{DSCnested}.   
It is clear that again the strong interference between the impurity 
wavefunctions survive the filtering effect and pose new
constraints on the potential scattering scenario versus more strongly
correlated models\cite{vojta}. We note that despite the very different
band structure used to calculate the LDOS in Fig. \ref{DSCnested} and Fig. \ref{DSCnotnested}, the
overall evolution of the resonances is quite similar except that the
apparent resonances are shifted to higher energies for the $t$-$t'$ band.
As mentioned above, it has been previously suggested that the nested ($t$-$t'$) band is 
appropriate for the DDW (DSC) state\cite{morr}. In that case we need
compare Fig. \ref{DDWnested} and Fig. \ref{DSCnotnested}. As opposed to the single impurity LDOS, the
configuration in Fig. \ref{DDWnested}b and Fig. \ref{DSCnotnested}b
again allows one to distinguish the DDW and DSC states by the number
of resonance peaks. This is contrary to Fig. \ref{DDWnested}a and
Fig. \ref{DSCnotnested}a which are remarkably similar.\\ 
\begin{figure}
\centerline{\epsfxsize=\linewidth\epsfbox{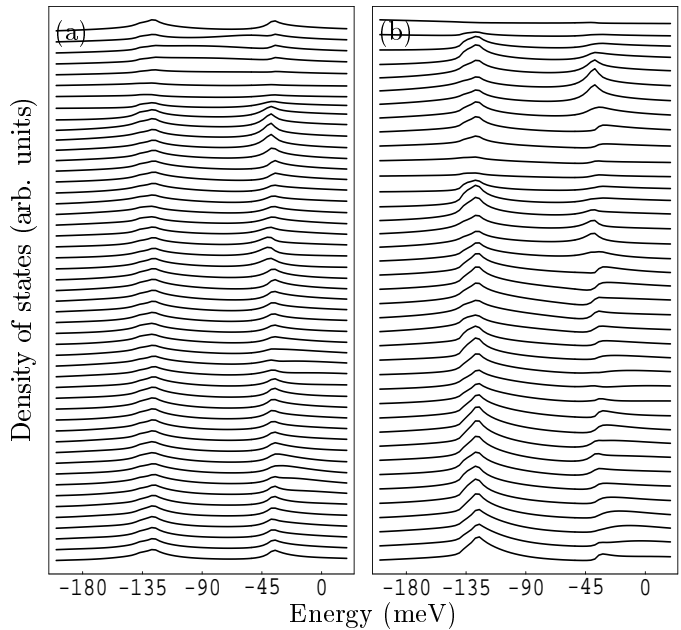}}
\caption{\label{DDWnotnested}Same as Fig. \ref{DSCnotnested} but for
the d-density wave state. Note the energy range. The peaks
evident in these scans are the shifted DDW gap edges (not impurity resonances).}
\end{figure}
\noindent In the DDW phase we know from the single impurity case that the 
current choice of band parameters leads to strongly
overdamped impurity resonances (Fig. \ref{1impDDW}c). 
However, for completeness we show the calculated STM scans
in Fig. \ref{DDWnotnested}. As expected the quantum interference is
weak and causes only minor changes in the DDW gap edges. Furthermore, 
the LDOS shown in Fig. \ref{DDWnotnested} changes only slightly upon
varying $U$ or the impurity positions.
\section{conclusion}
In summary we have shown that a systematic STM study around two nonmagnetic
impurities can clearly distinguish the DSC and DDW phases. In
particular, we suggest to perform STM scans with the positions of the
impurities being invariant under a mirror reflection through the scan
line. Even for the nested band, where the clean and the
single impurity LDOS are not a good probe for the underlying state,
this situation provides a robust test for DSC versus DDW order.\\
The impurities are modelled as potential scatterers and the results pose
further tests on this approach. An important question remains whether
phase fluctuations present about T$_c$ in the pseudo-gap state
are strong enough to wash out the interference patterns. This will be
discussed in a future publication\cite{andersen2}. It would also be interesting to study similar multiple impurity
interference effects within other pseudo-gap models and within
other proposed scenarios for the resonances around nonmagnetic
impurities in d-wave superconductors. In particular, within models
explaining the single impurity LDOS as a Kondo resonance arising from
a confined spinon\cite{vojta,sachdev}, one may expect more novel changes as the distance
between two nonmagnetic impurities is decreased. This is because the
cost of frustrated dimers decrease in this limit making it unfavorable
to break another dimer, and hence no spin is expected near the
nonmagnetic impurities.\\

I acknowledge stimulating correspondence with P. Hedeg\aa rd,
J. Paaske, S. Sachdev, and J.-X. Zhu.
This work is supported by the Danish Technical Research Council via
the Framework Programme on Superconductivity.

\end{multicols}

\begin{references}
\bibitem{comment2} This is also valid when the impurity positions
  coincide with the STM scan line as in Fig. \ref{DSCnested}b-\ref{DDWnotnested}b.
\end{references}
\end{document}